\documentclass[%
 preprint,
 amsmath,amssymb,
 aps,
prab,
]{revtex4-1}

\usepackage{graphicx}
\usepackage{dcolumn}
\usepackage{bm}
\usepackage{overpic}
\begin{document}

\title{Scissor-cross ionization injection in laser wakefield accelerators}

\author{Jia Wang}
\author{Ming Zeng}%
 \email[Author of correspondence: ]{zengming@ihep.ac.cn}
\author{Xiaoning Wang}
\author{Dazhang Li}
 \email[Author of correspondence: ]{lidz@ihep.ac.cn}
\author{Jie Gao}
\affiliation{Institute of High Engergy Physics, Chinese Academy of Sciences, 100049 Beijing, China}%
\affiliation{University of Chinese Academy of Sciences, 100049 Beijing, China}

\today
\begin{abstract}
We propose to use a frequency doubled pulse colliding with the driving pulse at an acute angle to trigger ionization injection in a laser wakefield accelerator. This scheme effectively reduces the duration that injection occurs, thus high injection quality is obtained. Three-dimensional particle-in-cell simulations show that electron beams with energy of $\sim 500\ \rm MeV$, charge of $\sim 40\ \rm pC$, energy spread of $\sim1\%$ and normalized emittance of a few millimeter milliradian can be produced by $\sim 100\ \rm TW$ laser pulses. By adjusting the angle between the two pulses, the intensity of the trigger pulse and the gas dope ratio, the charge and energy spread of the electron beam can be controlled.
\end{abstract}

\maketitle
The laser wake field accelerator (LWFA) proposed by Tajima and Dawson has attracted many attentions due to its orders-of-magnitude higher acceleration gradient than that of the conventional radio-frequency accelerators~\cite{TTajimaPRL1979}.
Great breakthroughs have been made in the past few years. For example, an electron beam with energy of 7.8 GeV has been generated in 20 cm~\cite{AJGonsalvesPRL2019}. The electron beams with energy spread in the sub-percentage level have been produced using density-tailored plasma~\cite{LTKePRL2021}. The 24-hour stable LWFA has been achieved by decoding sources of energy drift and jitter~\cite{ARMaierPRX2020}. Improving the output beam quality parameters including the energy spread, the beam charge, the emittance, the energy stability and so on has been a long-term goal in this society for the high-demanding applications such as plasma based light sources and colliders~\cite{WLeemansPT2009, EuPRAXIACDR2020, WTWangNature2021, MZengNJP2021}.

To optimize the output beam quality, many controlled injection schemes have been proposed such as pulse collision injections~\cite{FubGPRE2004,DavPRL2009,GloPRL2018}, density gradient injections~\cite{BranPOF2008,GedPRL2008,SchPRAB2010}, ponderomotive injections~\cite{UmsPRL1996,MZengNJP2020}, external magnetic field injections~\cite{VieJPRL2011,VieJPPCF2012,BuSVPOP2013,RassPOP2015}, ionization injections~\cite{ChenJAP2006,ChenPOP2012,McGPRL2010,PaKAPRL2010} and so on.
The ionization injection is to release electrons inside the pseudo-potential well of a wakefield by high-order ionization of the dopant species (high-Z elements such as Nitrogen, Oxygen, Neon or Argon). The injection amount can be adjusted by changing the density ratio of the dopant to the background plasma which is pre-ionized from low-ionization-threshold species such as Hydrogen and Helium. The ionization injection has the advantage of high reproducibility, but its energy spread is usually large. To reduce the energy spread, one may use the self-dechirping effect~\cite{LTKePRL2021, WTWangPRL2016}, and/or reduce the electrons injection length~\cite{MZengPOP2014, MMirzaieSrep2015}. For example, an electron beam with slice energy spread of $13\ \rm keV$ and charge of $0.4\ \rm pC$ can be produced by ionization injection of counter-propagating laser pulse~\cite{WanYPPCF2016}. An electron beam with slice energy spread of $12\ \rm keV$ and charge of $5\ \rm pC$ can be produced by two colliding lasers propagating in the transverse direction~\cite{LiFPRL2013}. The scheme of beat frequency ionization injection using frequency tripled laser was proposed for generating low energy spread electron beams~\cite{MZengPRL2015, MZengPOP2016}. However, this scheme has experimental difficulty due to lacking of high-efficiency frequency tripler.

\begin{figure}
\includegraphics[width=0.5\linewidth]{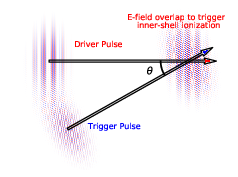}
\caption{\label{fig:1} Illustration of scissor-cross ionization injection. When the driver pulse and trigger pulse overlap in the plasma, a strong transient electric field is generated, which ionizes the inner shell electrons of the dopant atoms and produces ionization injection. 
}
\end{figure}

In this work, we propose a new scheme to trigger electron ionization injection by a frequency doubled pulse colliding with the driver pulse at an acute angle as illustrated in Fig.~\ref{fig:1}. A driver laser pulse drives a plasma wake (not shown in the figure) and collides with a trigger laser pulse at an angle $\theta$. During the collision, a higher superimposed electric field is generated, triggering the ionization of electrons of the inner shell of the dopant species. The driving laser itself can not ionize the inner shell of the dopant. The ionization injection only occurs when two laser pulses overlap, thus the injection is localized and the energy spread of the produced electron beam is limited. The frequency-doubled trigger laser has largely reduced ponderomotive force compared to a fundamental-frequency trigger laser with the same ionization electric field strength, thus its disturbance to the main wakefield is limited.
The trajectories of the driver and trigger lasers are similar to the two blades of a scissor, thus we call this scheme the scissor-cross ionization injection.

\begin{figure}
    \centering
    \begin{overpic}[width=0.72\textwidth]{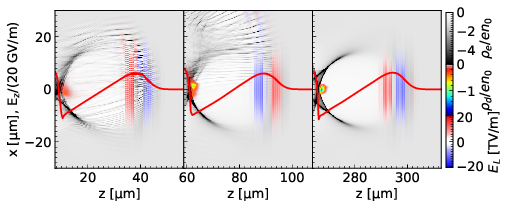}
        \put(13,10){(a)}
        \put(38,10){(b)}
        \put(62,10){(c)}
    \end{overpic}
    \caption{\label{fig:2} Snapshots of a 2D PIC simulation using the scissor-cross ionization injection scheme, (a) at the instant of time that the driver and trigger pulses overlap and the inner-shell ionization of the dopant occurs, (b) after the injection occurs, and (c) at a certain acceleration distance. The red curves show the line-out of the axial longitudinal electric field $E_z$ in unit of $20\ \rm GV/m$.
}
\end{figure}

The snapshots of the example two-dimensional (2D) particle-in-cell (PIC) simulation using the code WarpX~\cite{JLVayPOP2021} are shown in Fig.~\ref{fig:2}. In the plots, $n_0$ is the unperturbed plasma density, $\rho_e$ is the density of background electrons, $\rho_d$ is the density of electrons ionized from the dopant species, $E_L$ is the electric field of the laser pulses, and $e$ is the elementary charge. In the simulation, the background plasma is fully pre-ionized gas (e.g.\ hydrogen or helium) and the dopant is neon pre-ionized to +8 charge state. Because the ionization threshold of $\rm Ne^{8+}$ is $\sim 16.5\ \rm TV/m$, for the driver pulse with normalized vector potential amplitude $a_0<4$, it can be guaranteed that $\rm Ne^{8+}$ is not further ionized by the driver pulse, where $a_0 \approx 8.5\times 10^{-10} \lambda\left[{\rm \mu m}\right]\sqrt{I_0\left[{\rm W/cm^2}\right]} = E_0\left[{\rm TV/m}\right]\cdot \lambda\left[{\rm \mu m}\right]/3.2$ is the normalized vector potential amplitude, $I_0$ is the laser intensity and $E_0$ is the peak electric field strength of the laser. Meanwhile, by choosing $a_0>3$, we can ensure the self-guiding of the driver laser pulse for a sufficient longer electron acceleration distance without using a parabolic plasma channel~\cite{WLuPRAB2007,CBenedettiPOP2012}.
Practically, the driver pulse has normalized vector potential amplitude of $a_0=3.24$ and wavelength of $800\ \rm nm$, while the trigger pulse has normalized vector potential amplitude of $a_1=1.62$ and wavelength of $400\ \rm nm$. Both the two pulses have spot radius of $r_0 = r_1 = 15\ \rm \mu m$ and pulse duration of $30\ \rm fs$. Their focusing is synchronized spatially and temporally to trigger the inner-shell ionization of the dopant species. After the intersection finishes, the ionization injection does not occur anymore, because the electric field strength cannot reach the inner-shell ionization threshold of the dopant species even if the self-focusing of the driver laser pulse occurs. That is to say, ionization injection only occurs when two laser beams overlap, which ensures a limited injection distance.

\begin{figure}
\centering
    \begin{overpic}[width=0.98\textwidth]{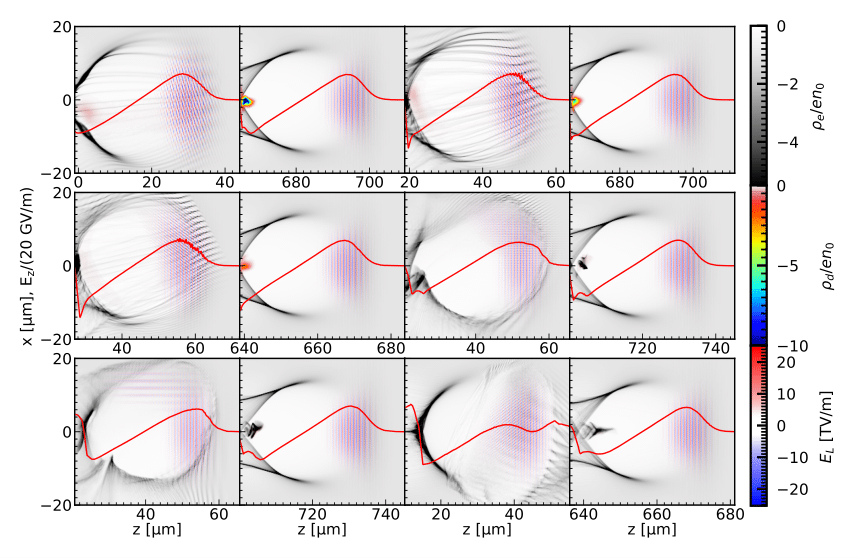}
        \put(22,58){(a.1)}
        \put(41,58){(a.2)}
        \put(60,58){(b.1)}
        \put(78.8,58){(b.2)}
        \put(22,39){(c.1)}
        \put(41,39){(c.2)}
        \put(60,39){(d.1)}
        \put(78.8,39){(d.2)}
        \put(22,20){(e.1)}
        \put(41,20){(e.2)}
        \put(60.5,20){(f.1)}
        \put(79.5,20){(f.2)}
    \end{overpic}
    \caption{\label{fig:3} Snapshots of simulations with the collision angles (a.1, 2) $\theta=10^\circ$, (b.1, 2) $\theta=20^\circ$, (c.1, 2) $\theta=30^\circ$, (d.1, 2) $\theta=60^\circ$, (e.1, 2) $\theta=90^\circ$ and (f.1, 2) $\theta=150^\circ$. The snapshots when the two pulses collide are (x.1) and the snapshots after certain acceleration distances are (x.2), where x stands for the letters from a to f. The line-out of the axial longitudinal electric field $E_z$ is plotted as red curves in the subplots.
}
\end{figure}

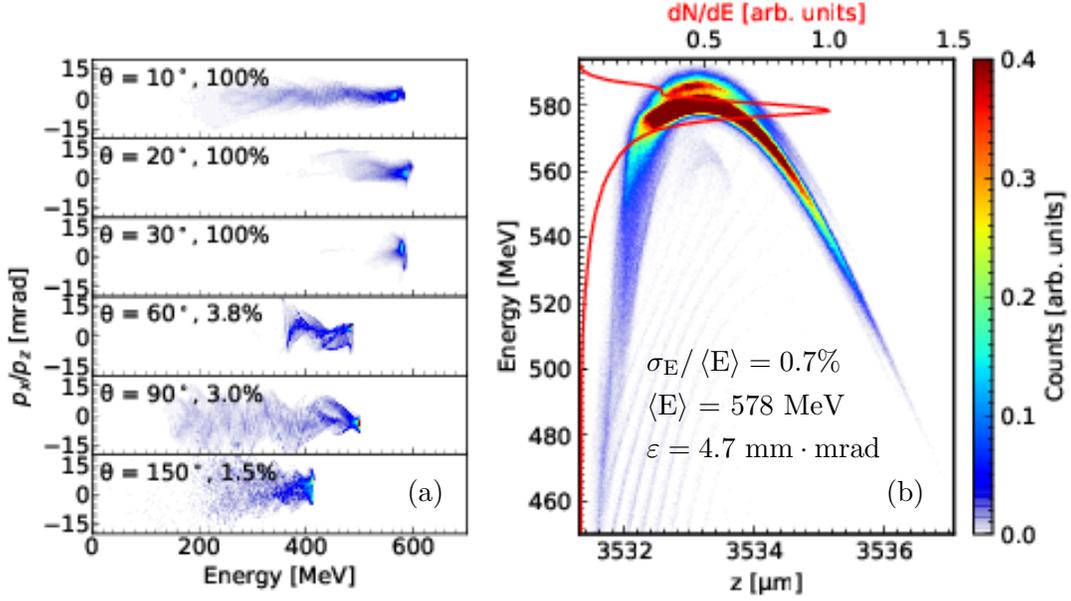
\begin{figure}
\centering
    \begin{overpic}[width=0.88\textwidth]{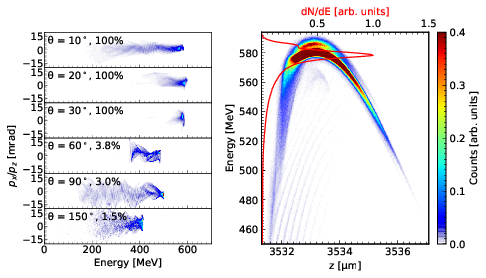}
        \put(38,10){(a)}
        \put(82,10){(b)}
        \put(60,22){\small $\rm \sigma_E/\left<E\right>=0.7\%$}
        \put(60,18){\small $\rm \left<E\right>$ = 578\ \rm MeV}
        \put(60,14){\small $\varepsilon=4.7\ \rm mm\cdot mrad$}
    \end{overpic}
    \caption{\label{fig:4} (a) The energy and angular distribution of electron bunches with different collision angles $\theta$. The percentages in the legend show the ratios of the charge from the inner-shell of the dopant species (i.e.\ from ionization injection) to the total injected charge. The ratio of transverse momentum to longitudinal momentum $p_x/p_z$ represents the divergence angle of the beam. (b) The phase space distribution of the electron beam at the acceleration distance of $3.5 \ \rm mm$ for $\theta=30^\circ$. The beam quality parameters are written in the plot: the mean energy is $\rm \left<E\right> = 578\ MeV$, the RMS energy spread is $\rm \sigma_E/\left<E\right>=0.7\%$, and the transverse normalized emittance is $\varepsilon=4.7\ \rm mm\cdot mrad$.
}
\end{figure}

To study the effect of different collision angles, we have performed simulations with $\theta$ varying from $10^\circ$ to $150^\circ$. The snapshots and the quality of the injected beams are shown in Fig.~\ref{fig:3} and Fig.~\ref{fig:4}, respectively. These simulations use pure Neon element, which provides both the pre-ionized outer shell to form the background plasma and the bounded inner shell for ionization injection. As one can see, the beam quality changes with $\theta$. For $\theta\lesssim 30^\circ$, the disturbance of the trigger pulse to the main wakefield is small, but the length of the region that the two pulses overlap, and thus the injection length which can be estimated by $r_0/\tan \theta$, is relatively large. For $\theta\gtrsim 60^\circ$, the injection length can be $\sim 10\ \rm \mu m$, but there is a significant disturbance of the trigger pulse to the main wakefield which degrades the injected beam quality. Moreover, for $\theta \lesssim 30^\circ$, the injection mechanism is purely ionization injection, while for $\theta \gtrsim 60^\circ$, the injection mechanism is gradually switched to colliding pulse injection as shown in Fig.~\ref{fig:4} (a). The phase space distribution of the output beam for the case $\theta = 30^\circ$ is plotted in Fig.~\ref{fig:4} (b), which shows a beam with a mean energy of $578\ \rm MeV$, a root-mean-square (RMS) energy spread of $0.7\%$, and a normalized emittance of $4.7\ \rm mm\cdot mrad$. 

\begin{figure}
\centering
    \begin{overpic}[width=0.78\textwidth]{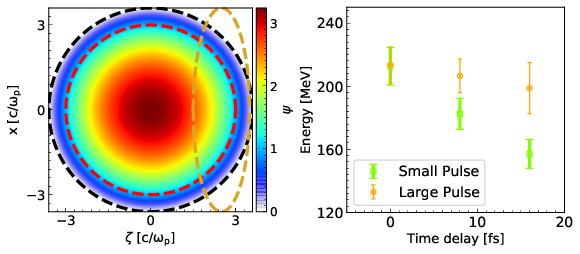}
        \put(10,39){(a)}
        \put(61,39){(b)}
    \end{overpic}
    \caption{\label{fig:5} (a) Illustration of pseudo-potential distribution in the co-moving frame of the driver laser. The driver laser has $a_0=3.24$ and $r_0=15\ \rm \mu m$, and the plasma density is $1.63\times10^{18}\ \rm cm^{-3}$. The black dashed curve represents the boundary of the bubble and the dashed red curve encircles the region of inner-shell ionization in which the trapping condition $\Delta\psi\leq -1$ is possible to be satisfied. The golden ellipse represents the profile of the driver laser which is propagating towards the $+\zeta$ direction. (b) The mean energy (dots) and energy spread (vertical error-bars) vs.\ time delay of the trigger pulse relative to the driver for two cases of the trigger sizes obtained by 2D PIC simulations. $a_1$ is fixed to 1.62 for the trigger laser, the dopant $\rm Ne^{8+}$ density is fixed to 3\% of the pre-ionized plasma density, the acceleration distance is fixed to $1\ \rm mm$ and the collision angle is fixed to $\theta=8^\circ$. The orange symbols represent the case with a larger trigger spot size $r_1=15\ \rm \mu m$ and a longer trigger pulse duration $\tau_1=30\ \rm fs$, and the green symbols represent the case with a smaller trigger spot size $r_1=3 \ \rm \mu m$ and a shorter trigger pulse duration $\tau_1=6\ \rm fs$.}
\end{figure}

Obviously, the strength of the superimposed electric field of the driver and trigger does not depend on their sizes. However, the sizes of the pulses determine the injection quantity and also the injection difficulty in a real experiment. We discuss the effect of changing the trigger size in the following. Assume the wakefield is a spherical bubble, the pseudo-potential in the bubble can be estimated by $\psi \approx \omega_p^2(r^{2}_b(\zeta)-r^2)/4c^2$, where $r_b$ is the bubble radius, $r$ is the distance to the central axis, $\zeta=z-ct$ is the co-moving coordinate, $c$ is the speed of light in vacuum, $\omega_p=c\sqrt{4\pi r_e n_0}$ is the plasma frequency and $r_e$ is the classical electron radius~\cite{IKostyukovPOP2004, LuWPOP2006}. The condition of trapping for an electron is $\Delta\psi\lessapprox -1$ if the electron has negligible initial momentum~\cite{MZengNJP2020}. In a plasma with the density $n_0 = 1.63\times10^{18}\ \rm cm^{-3}$, the laser with $a_0=3.24$ and the matched spot size $r_0=3.6c/\omega_p=15\ \rm \mu m$ drives a wakefield with its distribution of pseudo-potential illustrated in Fig.~\ref{fig:5} (a). The electrons been ionized in the dashed red circle satisfy the trapping condition and can be captured by the wakefield. A larger spot size and longer pulse duration of the trigger can decrease the influence of the time delay jitter of the two pulses to the output electron beam energy as one can see in Fig.~\ref{fig:5} (b). This is because the time delay jitter of a smaller trigger laser more significantly influences the injection phase of the electron beam, which determines the acceleration field strength exerted on the beam.

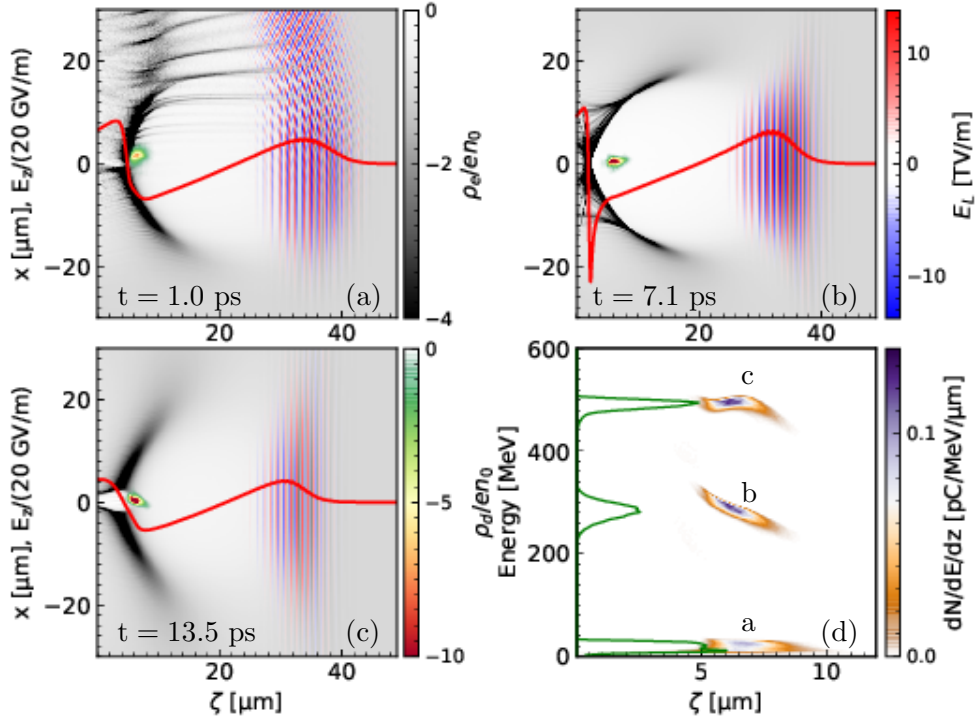
\begin{figure}
\centering
    \begin{overpic}[width=0.8\textwidth]{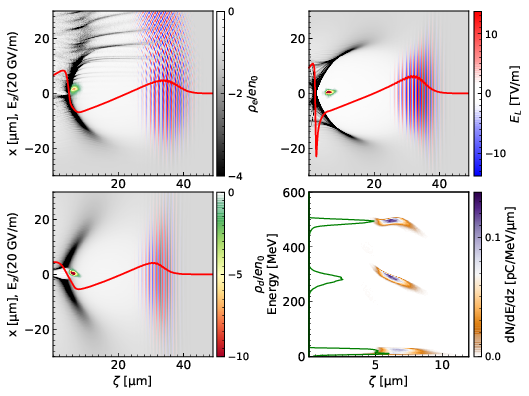}
        \put(35,43){(a)}
        \put(12,43){$\rm t=1.0\ ps$}
        \put(83,43){(b)}
        \put(60,43){$\rm t=7.1\ ps$}
        \put(35,9){(c)}
        \put(12,9){$\rm t=13.5\ ps$}
        \put(83,9){(d)}
        \put(75,10){a}
        \put(75,23){b}
        \put(75,35){c}
    \end{overpic}
    \caption{\label{fig:6} Plots of a 3D PIC simulation. The driver pulse has $a_0=2.74$, the trigger pulse has $a_1=1.37$, and the collision angle is $\theta=8^\circ $. The snapshots are taken (a) short after colliding, (b) at about $2\ \rm mm$ acceleration distance and (c) at about $4\ \rm mm$ acceleration distance. (d) The phase space distribution and energy spectrum (green curves) of the trapped electron beam at the time corresponding to the above three snapshots.}
\end{figure}

We have also performed fully three-dimensional (3D) simulations to verify our scheme. Because the self-focusing effect of the laser in plasmas is stronger in a 3D case than in a 2D case, the former 2D simulation parameters are not suitable for the 3D simulations.
The new parameters for a 3D simulation are the following. The driver and trigger pulses have normalized vector potential amplitude of $a_0=2.74$ and $a_1=1.37$, respectively, and they have the same focal waist radius of $r_0=r_1=20\ \rm \mu m$. They are both polarized in the $y$ direction and collide at $\theta=8^\circ$. We do not choose $\theta=30^\circ$ as in Fig.~\ref{fig:4}(b) because a larger $\theta$ significantly increases the computational cost to an unaffordable level. The pre-ionized background plasma density is $n_0=1.36\times10^{18}\ \rm cm^{-3}$ and the density of $\rm Ne^{8+}$ is $n_0/8$. The simulation shows that the maximum electric field strength after occurring of self-focusing is $15\ \rm TV/m$ which is sufficiently smaller than the ionization threshold of $\rm Ne^{8+}$ (which is around $17\ \rm TV/m$), thus the driver pulse itself does not trigger ionization injection.
The simulation has the moving window size of ($100\ \rm \mu m$, $100\ \rm \mu m$, $50\ \rm \mu m$) and the cell number of (768, 128, 2432) for ($x$, $y$, $z$) directions, respectively. 
The results are shown in Fig.~\ref{fig:6}. The electron beam is injected by our scheme at $t=1.0\ \rm ps$ with a small absolute energy spread (Fig.~\ref{fig:6} (a)) and mainly experiencing positive chirping during the acceleration (Fig.~\ref{fig:6} (b)). At the acceleration distance of about $4\ \rm mm$ (Fig.~\ref{fig:6} (c)), the electron beam enters the negative chirping region due to the weakening of the driver, thus the initial small absolute energy spread is retrieved (phase space is shown in Fig.~\ref{fig:6} (d)). The beam after the acceleration distance of $4\ \rm mm$ has the charge of $Q=40\ \rm pC$, the peak current of $I_{\rm peak}=6.22\ \rm kA$, the mean energy of $\rm \left<E\right>=500\ \rm MeV$, the RMS energy spread of $\rm \sigma_E/\left<E\right>=1.6\%$, and the normalized emittance of $\varepsilon_x=1.11\ \rm mm\cdot mrad$, $\varepsilon_y=7.84\ \rm mm\cdot mrad$ for the two transverse directions, respectively.

In conclusion, we have introduced the scissor-cross ionization injection scheme in LWFA, which uses the intense electric field generated at the moment when the trigger pulse overlaps with the driver laser pulse to ionize the inner shell electrons of the dopant species and to trigger the ionization injection. Both 2D and 3D simulations show that this scheme produces high quality electron beams with the energy spread of the order of 1\%. The ionization injection is limited to a small region with the length of $\sim r_0/\tan\theta$ which is typically $\lesssim 100\ \rm \mu m$, where $r_0$ is the spot size of the driver pulse and $\theta$ is the collision angle. It is theoretically possible to further decrease the energy spread to per-mille level by slightly increasing $\theta$ and/or by increasing the acceleration distance with a preformed plasma channel.
In experimental implementation of this scheme, the trigger pulse can be replaced by a laser with any frequency, as long as the superimposed electric field exceeds the inner-shell ionization threshold of the dopant species when the two pulses collide. However, with a certain peak electric field strength, a lower-frequency laser has larger ponderomotive force which may degrade the injection quality. We mainly use the frequency doubled trigger in our discussion, for its reasonable experimental difficulty and relatively small ponderomotive disturbance to the main wakefield.

\acknowledgements
This work is supported by the Research Foundation of Institute of High Energy Physics, Chinese Academy of Sciences (Grant No.\ E05153U1, No.\ E15453U2, Y9545160U2 and Y9291305U2). This research used the open-source particle-in-cell code WarpX primarily funded by the US DOE Exascale Computing Project. We acknowledge all WarpX contributors.

\nocite{*} 
\providecommand{\noopsort}[1]{}\providecommand{\singleletter}[1]{#1}%
\end{document}